\documentclass[11pt]{article}
\usepackage{amsmath,amssymb,amsfonts}
\usepackage[dvips]{graphicx}
\usepackage{epsfig}
\usepackage{cite}

\makeatletter \@addtoreset{equation}{section} \makeatother
\makeatletter \@addtoreset{figure}{section} \makeatother

\def\IC{\mathbb{C}}

\def\IR{\mathbb{R}}\def\IZ{\mathbb{Z}}

\def\CA{{\cal A}}\def\CD{{\cal D}}
\def\CE{{\cal E}}

\def\CL{{\cal L}}\def\CM{{\cal M}}
\def\CN{{\cal N}}\def\CO{{\cal O}}\def\CP{{\cal P}}

\def\CZ{{\cal Z}}

\def\a{\alpha}\def\b{\beta}\def\g{\gamma}
\def\d{\delta}\def\e{\epsilon}

\def\m{\mu}\def\n{\nu}
\def\r{\rho}\def\s{\sigma}
\def\t{\tau}
\def\w{\omega}
\def\L{\Lambda}
\def\O{\Omega}

\def\p{\partial}

\begin{document}

\begin{titlepage}
\vfill

\begin{flushright}
{\tt\normalsize KIAS-P10018}\\
\end{flushright}
\vfill

\begin{center}
{\Large{\bf  1/2-BPS Wilson Loops and Vortices in ABJM Model
\mathversion{normal}
\par}
    }\vspace{20mm}
{ Ki-Myeong Lee, and Sungjay Lee}\\[5mm]
{
Korea Institute for Advanced Study, Seoul 130-722, Korea
}\\ [5mm]
{\tt   klee@kias.re.kr,sjlee@kias.re.kr }
\vfill
\end{center}

\begin{abstract}
We explore the low-energy dynamics of 1/2-BPS heavy particles
coupled to the ABJM model via the Higgsing of M2-branes, with
focus on physical understanding of the recently discovered 1/2-BPS
Wilson loop operators. The low-energy theory of 1/2-BPS heavy particles
turns out to have the $U(N|N)$ supergauge symmetry,
which explains the novel structure of the 1/2-BPS Wilson loop operator as
a holonomy of a $U(N|N)$ superconnection. We show that the supersymmetric
transformation of the Wilson loop operator can be identified as a
fermionic supergauge transformation, which leads to their invariance under
half of the supersymmetry. We also  argue that 1/2-BPS Wilson loop operators  appear as
1/2-BPS vortices with vorticity $1/k$. Such a vortex   can be naturally interpreted as a membrane
wrapping the $\IZ_k$ cycle once, or type IIA fundamental string.
\end{abstract}
\vfill

\end{titlepage}

\renewcommand{\thefootnote}{\#\arabic{footnote}}
\setcounter{footnote}{0}

\section{Introduction and Conclusion}

It has been recently proposed by Aharony, Bergman, Jafferis and Maldacena
that the $\CN=6$ superconformal Chern-Simons theory with gauge group
$U(N)\times U(N)$  describes the low-energy dynamics on N M2-branes
at the tip of the orbifold space $\mathbb{C}^4/Z_k$ \cite{Aharony:2008ug}.
The gravity dual of the theory is either M-theory on AdS$_4\times S^7/\mathbb{Z}_k$ background,
or type IIA string theory on AdS$_4\times \mathbb{CP}^3$, depending on
the range of the Chern-Simons level $k$.

A set of physical observables in the conformal theories is
spanned by correlation functions of gauge invariant operators.
As important and interesting observables, the BPS Wilson loop
operators have gotten  great attention due to their clear identification
as macroscopic strings in the gravity dual \cite{Rey:1998ik,Maldacena:1998im}.
In the $\CN=6$ Chern-Simons theory, the half BPS Wilson line operator corresponding
to the most symmetric string configuration has  remained unidentified
for a while, despite the immediate initial discoveries of 1/6-BPS
Wilson line operator \cite{Drukker:2008zx,Chen:2008bp,Rey:2008bh}
and 1/2-BPS vortices \cite{Drukker:2008jm}.

The very 1/2-BPS Wilson line operator of $\CN=6$ Chern-Simons theory
has been constructed recently in \cite{Drukker:2009hy}. It turns out that the operator
has some interesting and novel features. More precisely, it takes
the form as a holonomy of the superconnection in
the super Lie group $U(N|N)$ which is related to  the $U(N)\times U(N)$ ABJM model.
The superconnection also involves certain constant spinors.
Using the localization technique as in \cite{Pestun:2007rz},
one can even compute the vacuum expectation value of the 1/2 BPS Wilson
line operator exactly \cite{Kapustin:2009kz,Suyama:2009pd,Marino:2009jd}.

In this work, we focus on how to understand the above
peculiar structure of the 1/2 BPS Wilson line operator via
a systematic and physical procedure, so-called Higgsing.

One possible interpretation of the Wilson line operator is
an insertion of an external charged particle into the given system.
The Wilson line operator appears as how the wave-function of
the external particle  evolves under its interactions
with the given system. In order to study the BPS Wilson line operator,
it is therefore essential how to introduce such very heavy particles
to the system in a supersymmetric fashion. It is simply provided by
the suitable Higgsing procedure.

We systematically explore the low-energy dynamics of
1/2-BPS very massive particles in the Coulomb phase of the M2-brane theory
that describes a separation of a single M2-brane far away
from the rest M2-branes placed near the orbifold singularity.
In the infinite separation  limit, these infinitely massive particles
obviously provide the external source as the 1/2 BPS Wilson line operator.
In this paper, we pay attention in particular how the fermion
fields can affect the time-evolution of the external particles,
which leads to a physical explanation of the super Lie algebraic
structure of the Wilson line operator. Although the full ABJM model itself
does not have the $U(N|N)$ supergauge symmetry,
the low-energy theory of these external 1/2-BPS
particles coupled to ABJM model turns out to respect the supergauge symmetry.

We also show in addition that the supersymmetry transformation
of 1/2-BPS Wilson line operator can be regarded as   a
gauge transformation of the superconnection with fermionic
super Lie algebra elements. It immediately implies the
SUSY invariance of loop operators under trace or supertrace
depending on whether the fermionic field is anti-periodic or periodic.
For the abelian ABJM model, the insertion of the 1/2-BPS
Wilson line operator creates a fundamental 1/2-BPS vortex,
which can be described as an M2-brane spike wrapping
the $S^1$ fibre of $S^7/\mathbb{Z}_k$.
We also study several properties of the vortices in relation
to the 1/2-BPS Wilson line operator.

Our analysis can be also applied to less supersymmetric
cases, either Wilson line operators or theories.
For examples, for 1/6-BPS Wilson lines, one can quickly
show that all interactions of 1/6-BPS external particles
that could deliver fermionic contributions to Wilson lines
are averaged out in the infinite mass limit, due to their highly oscillatory
behaviors. The absence of those interactions explains the usual expression
for 1/6-BPS Wilson line operator.
One can also show that there is 2/5-BPS Wilson line operators
in $\CN=5$ Chern-Simons matter theories \cite{Hosomichi:2008jb,Aharony:2008gk},
similar to 1/2-BPS Wilson line operators
in the ABJM model after suitable changes in gauge group representation
and additional reality conditions. The low-energy dynamics of
corresponding external particles are again
expected to have the $OSp(2N|2N)$ supergauge symmetry.

The super Lie algebraic structure of 1/2-BPS Wilson
line operators however severely restricts their possible representations under
the gauge group, since the fermionic components should transform
as bi-fundamental representations of $U(N)\times U(N)$.
It would be interesting how to circumvent this difficulty,
or understand the obstruction in the membrane picture.
The insertion of the Wilson-loop affects the system to break the scale
symmetry, based on the physics of 1/2-BPS vortex in abelian theories. We
expect that this could persist in the weak coupling or large $k$ limit.
It would be also interesting to understand this physics in detail.

This paper is organized as follows.
In Section 2, we briefly review the ABJM model and discuss the
mass spectrum in the Higgsing procedure. The massive modes can be
identified as macroscopic membranes, interpolating the separated
M2-branes and wrapping the $S^1$ fibre of orbifold $\mathbb{C}^4/\mathbb{Z}_k$ once.
We study in Section 3 the Higgsing procedure in more details
to provide external particles to the ABJM model in a supersymmetric fashion.
We discuss first how to read off the dynamical modes from the off-diagonal
massive fields in the infinite mass limit, which parallels to
what we perform in non-relativistic limit. We then present some delicate
points in obtaining the low-energy theory of 1/2-BPS heavy particles,
which originates from non-trivial interactions between the heavy
particles and ABJM fields.
In section 4, we show that low-energy theory of external 1/2-BPS particles
coupled to the ABJM model preserves the $U(N|N)$ supergauge symmetry.  It
explains the physical origin of the novel structure of the 1/2-BPS Wilson
line operators.  We give an alternative proof on
the invariance of 1/2-BPS Wilson line operator under the half of
supersymmetry in relation to the supergauge transformation.
We finally study in Section 5 the interesting relations between
1/2-BPS vortices of vorticity $1/k$ and 1/2-BPS Wilson line.
We briefly discuss in Appendix the infinite mass limit in the free
field theories, with care given to dynamical modes which
can survive in the low-energy theory.

\section{Preliminaries}

\subsection{Short review on ABJM model}

Let us start with a short description on the  ABJM model
\cite{Aharony:2008ug}, believed to describe the dynamics of
multiple M2-branes probing a  orbifold  $\IC^4/\IZ_k$.
It is the ${\cal N}=6$ supersymmetric Chern-Simons matter theory
with the gauge group $G=U(N ) \times U(N )$. The gauge fields are now
denoted by $A_\mu$ and $\tilde A_\mu$ with the Chern-Simons levels $(k,-k)$.
The matter fields are composed of four complex scalars
$Z_\a$ ($\a=1,2,3,4$) and four three-dimensional spinors
$\Psi^\a$, both of which transform under the gauge symmetry
as bi-fundamental representations $({\bf N}, \bar{\bf N})$.
As well as the gauge symmetry, the present model also has additional global
$SU(4)_R$ symmetry, under which the scalars $Z_\alpha$ furnish the representation
${\bf 4}$ while the fermions $\Psi^\alpha$ furnish ${\bf \bar 4}$. Both $Z_\a, \Psi^\a$ carry an abelian charge which identifies particles and anti-particles.

The ABJM Lagrangian takes the following forms
\begin{eqnarray}
  {\cal L}={\cal L}_\text{CS} + {\cal L}_\text{kin} +
  {\cal L}_\text{Yukawa} + {\cal L}_\text{potential}\ .
\end{eqnarray}
The Chern-Simons terms and matter kinetic terms are
\begin{eqnarray}
  {\cal L}_\text{CS} + {\cal L}_\text{kin}
  &=& \frac{ k  }{4\pi }\e^{\mu\nu\rho}
  \text{Tr} \Big( A_\mu \partial_\nu A_\rho -  \frac{2i}{3}
  A_\mu A_\nu A_\rho  - \tilde{A}_\mu \partial_\nu \tilde{A}_\rho
  +  \frac{2i}{3}  \tilde{A}_\mu \tilde{A}_\nu \tilde{A}_\rho \Big)
  \nonumber \\ &&
  \hspace{0.3cm}- \text{Tr}  \left( D_\mu \bar{Z}^\a D^\mu Z_\a +
  i \bar{\Psi}_\a \g^\mu D_\mu \Psi^\a \right)\  ,
\end{eqnarray}
where
  $ D_\mu Z_\a =
\partial_\mu Z_\a - i A_\mu Z_\a + i Z_\a \tilde A_\mu$ and so on.
The Yukawa-like interactions are
\begin{eqnarray}
  {\cal L}_\text{Yukawa}  &=&
  \frac{2\pi i}{k }\text{Tr}
  \Big(  \bar{Z}^\a Z_\a \bar{\Psi}_\b \Psi^\b
  - Z_\a \bar{Z}^\a \Psi^\b \bar{\Psi}_\b
  +2 Z_\a \bar{Z}^\b \Psi^\a \bar{\Psi}_\b
  -2 \bar Z^\a Z_\b \bar{\Psi}_\a \Psi^\b \nonumber \\
  && \hspace{1.3cm}
  + \epsilon_{\a\b\g\d} \bar{Z}^\a \Psi^\b \bar{Z}^\g \Psi^\d
  -  \epsilon^{\a\b\g\d} Z_\a \bar{\Psi}_\b Z_\g \bar{\Psi}_\d
  \Big)\ .
\end{eqnarray}
The sextic scalar interactions are summarized simply as
\begin{eqnarray}
   {\cal L}_\text{potential} &=& -U= -
  \frac{4\pi^2}{3k^2  } \text{Tr}\Big(
  6 Z_\a \bar Z^\a Z_\b \bar{Z}^\g Z_\g \bar Z^\b
  - 4Z_\alpha \bar{Z}^\b Z_\g  \bar Z^\a Z_\b \bar Z^\g \nonumber \\
  && \hspace{1.9cm}
  - Z_\a \bar Z^\a Z_\b \bar Z^\b Z_\g \bar Z^\g
  - Z_\a\bar Z^\b Z_\b \bar Z^\g Z_\g \bar Z^\a  \Big)\ .
\end{eqnarray}
The positive definite bosonic potential $U$ can be expressed in terms of
third order polynomials $W$ and its hermitian conjugate $\bar W$:
\begin{eqnarray}
  U= \frac23 \text{Tr}\Big(W^\a_{\b\g} \bar{W}^{\b\g}_\a\Big) \ge 0
\end{eqnarray}
with
\begin{eqnarray}
  && W_{\b\g}^\a =-  \frac{\pi}{k  }
  \Big( 2Z_\b \bar Z^\a  Z_\g +\d^\a_\b
  (Z_\g \bar Z^\rho Z_\rho -   Z_\rho\bar Z^\rho Z_\g)
  \Big) - (\b \ \leftrightarrow \  \g)\ , \nonumber \\
  && \bar W^{\b\g}_\a = + \frac{\pi}{k }
  \Big(  2\bar Z^\b Z_\a  \bar Z^\g
  +\d_\a^\b (\bar Z^\g Z_\rho \bar Z^\rho-  \bar Z^\rho  Z_\rho \bar Z^\g)
  \Big)  - (\b \ \leftrightarrow \  \g)\ .
\end{eqnarray}

The ABJM model is invariant
under the ${\cal N}=6$ supersymmetry whose transformation rules
are summarized as
\begin{eqnarray}
  \d Z_\a  &=& i \xi_{\a \b} \Psi^\b\ ,  \qquad
  \d \Psi^\a  = -  \g^\mu \xi^{\a\b}  D_\mu Z_\b +W^\a_{\b\g} \xi^{\b\g}\ ,
  \nonumber \\
  \d \bar{Z}^\a &=& i\xi^{\a\b}\bar{\Psi}_\b\ , \qquad
  \d \bar{\Psi}_\a =  -  \g^\mu  \xi_{\a\b}   D_\mu  \bar{Z}^\b
  + \bar{W}^{\b\g}_\a \xi_{\b\g}\ , \nonumber \\
  \d A_\mu &=&  + \frac{2\pi }{k  }
  ( Z_\a\bar{\Psi}_\b \g_\mu \xi^{\a\b} + \Psi^\a \bar Z^\b  \g_\mu\xi_{\a\b})\ ,
  \nonumber \\
  \d \tilde{A}_\mu &=&
  - \frac{2\pi  }{k  } (\bar\Psi_\a Z_\b \g_\mu \xi^{\a\b}
  +\bar Z^\a\Psi^\b \g_\mu \xi_{\a\b})\  .
  \label{susytrans}
\end{eqnarray}
Moreover, this M2-brane theory is also invariant under the parity operation accompanied by
$$
Z_\a,\Psi^\a, A_\mu, \tilde{A}_\mu \ \
\leftrightarrow  \ \
\bar{Z}^\a , \bar{\Psi}_\a, \tilde{A}_\mu, A_\mu\ .
$$
Note that the supersymmetry transformation parameters $\xi^{\a\b}=-\xi^{\b\a}$
satisfy the reality condition
\begin{eqnarray}
  \xi_{\a\b} =(\xi^{\alpha\beta})^* =
  \frac12 \e_{\a\b\g\d} \xi^{\g\d}\ ,
  \label{12susy}
\end{eqnarray}
with the convention $\epsilon_{1234}=\epsilon^{1234}=1$. For later
convenience, one summarizes the equation of motions for gauge fields
\begin{eqnarray}
  \frac{k}{4\pi} \epsilon^{\mu\nu\rho} F_{\nu\rho}
  -i (Z_\a D^\mu \bar Z^\a - D^\mu Z_\a \bar Z^\a) &=& 0  \nonumber \\
  -\frac{k}{4\pi} \epsilon^{\mu\nu\rho} \tilde F_{\nu\rho}
  -i (\bar Z^\a D^\mu  Z_\a - D^\mu \bar Z^\a Z_\a) &=& 0\ .
\end{eqnarray}

Let us now examine the vacuum moduli space of the present model at
the classical level, i.e., solutions of $U(Z_\a,\bar Z^{ \b}) =0$ up to gauge
transformations. It leads to the equation for its minima
\begin{equation}\label{vac}
 Z_\alpha \bar Z^\beta Z_\gamma = Z_\gamma \bar Z^\beta Z_\alpha\,,~~~
 \bar Z^\alpha Z_\beta \bar Z^\gamma = \bar Z^\gamma Z_\beta \bar Z^\alpha\,.
\end{equation}
This implies that the hermitian matrices $Z_\alpha\bar Z^\beta$
commute with each other, and similarly for $\bar Z^\alpha Z_\beta$.
The vacuum solutions are thus
given by diagonal $Z_\alpha$ up to gauge equivalences,
\begin{eqnarray}
  Z_\a  = \text{diag}(z^1_\a, z^2_\a, .. , z^N_\a)\   \label{zdiag} .
\end{eqnarray}
On a generic point of the vacuum moduli space, the gauge group
$G=U(N)\times U(N)$ is spontaneously broken
down to $U(1)^N \subset U(N)_D$, diagonal part of $G$.

\paragraph{Convention} As a final comment, let us summarize our convention.
A natural choice for gamma matrices $\g^\mu$ is in the Majorana representation:
\begin{eqnarray}
  \g^0 = i\tau^2\ , \ \ \g^1=\tau^1\ , \ \ \g^2=\tau^3\ ,
  \ \ \g^{012}= {\bf 1}_2\ .
\end{eqnarray}
We define constant two-component spinors with definite
helicity $u_\pm$ as
\begin{eqnarray}\label{constspinor}
  u_\pm \equiv \frac{1}{\sqrt{2}}
  \left( \begin{array}{c} 1 \\  \mp i \end{array}\right)\ ,
  \ \  \bar u_\pm \equiv \frac{1}{\sqrt{2}} (1, \mp i)\ ,
\end{eqnarray}
which obviously satisfy the following relations
\begin{eqnarray}
  i\g^0 u_\pm = \pm u_\pm \ , \qquad
  \bar u_\pm  u_\mp =1\ , \qquad \bar u_+  u_+=0\ ,
\end{eqnarray}
and so on.

One can express the supersymmetry parameters $\xi^{\a\b}$ and Dirac spinors
in the helicity basis like below
\begin{eqnarray}
  \xi_{\a\b} &=&  u_+ \xi_{\a\b-} + u_- \xi_{\a\b+}\ ,
  \nonumber \\
  \Psi^\a &=&  u_+ \Psi^\a_-   +  u_-  \Psi^\a_+
\end{eqnarray}
where $\xi_{\a\b\pm}= [\bar u_\pm\xi_{\a\b}]= u_\pm^T \xi_{\a\b}$,
$\Psi^\a_{\pm}=  [\bar u_\pm\Psi^\a]= u_\pm^T\Psi^\a$.

\subsection{Higgsing and massive particles}

We now in turn discuss massive modes in the Higgsing procedure which
separates a single M2-brane apart from the rest N M2-branes at the
orbifold singularity. For concreteness, we first present the mass spectrum
at the generic point on the Coulomb branch of $U(2)\times U(2)$ ABJM model.
The mass formula for massive modes, applicable particularly in the Higgsing,
will be presented in order.

The generic vacuum of $U(2)\times U(2)$ ABJM model can be described as
\begin{eqnarray}
  \langle Z_\a \rangle = \text{diag} \big( u_\a ,v_\a \big)\ ,
\end{eqnarray}
where $u_\a$ and $v_\a$ denote the positions of two M2-branes in the
orbifold space $\mathbb{C}^4/\mathbb{Z}_k$. When $u_\a \neq v_\a$,
the linear fluctuation analysis tells us that the mass spectrum can
be summarized as
\begin{eqnarray}
  \left\{
    \begin{array}{ccc}
    \text{massless multiplet} & :& 16
    \ \text{scalar bosons}+ 16 \ \text{fermions} \\
    \text{massive multiplet}& : & 12
    \ \text{scalar bosons}+16\ \text{fermions}+ 4 \ \text{vector bosons}
    \end{array}
  \right.
\end{eqnarray}
The massive modes are made of a pair of 1/2 BPS massive vector multiplets of opposite parity as we will see.
The above massive modes arise from the off-diagonal elements of matter
fields, as usual. One can show that the perturbative mass $\mu$ of the massive
multiplet takes the following form \cite{Berenstein:2008dc}
\begin{eqnarray}\label{massformula1}
  \mu &=& \frac{2\pi}{|k|}\sqrt{(u_\a \bar u^{\a}  +   v_\a\bar v^\a )^2
  - 4|u_\a\bar v^\a|^2}\ .
\end{eqnarray}

For the Higgsing, we take the particular position of a single M2-brane
far away from the origin such that $v_\a=(0,0,0,v)$ and  for any $\a$  $|u_\a|\ll v $.
Then, the mass formula (\ref{massformula1}) becomes
\begin{eqnarray}\label{massformula2}
  \mu= \frac{2\pi}{k} \big(|u_1|^2   + |u_2|^2   + |u_3|^2   - |u_4|^2 \big)
  + \frac{2\pi}{k} |v|^2\ .
\end{eqnarray}
The quadratic dependence on the position parameters can be understood as
follows: the massive modes arise from an M2-brane which interpolates
two separated M2-branes and also wraps the 11-dimensional circle of
size proportional to `$\frac{2\pi}{k}\times\text{distance}$' \cite{Berenstein:2008dc}.
The signs in (\ref{massformula2}) imply that the distance between two M2-branes
increases along $z_1,z_2,z_3$ directions while decreases along $z_4$ direction, where
$z_\a \in \mathbb{C}^4/\mathbb{Z}_k$.

The presence of the W bosons and other massive particles
can be treated as source terms for scalars with some sign difference.
The end result becomes
\begin{eqnarray}
  \CL_{scalar} = - \frac{2\pi}{k}\big( |u_1|^2 + |u_2|^2 + |u_3|^2 - |u_4|^2\big)
  \delta^2 (z-z_p) \ ,
\end{eqnarray}
where $z_p$ denote the position of the source.
The energy contribution to the $u_1,u_2,u_3$ is increasing
and that to $u_4$ is decreasing.
Indeed the insertion of 1/2 BPS Wilson line to an abelian theory
leads to such scalar field source as we will see later in Section 5.

Let us finally consider the Higgsing for $N$ M2-branes, i.e., put
$N-1$ M2-branes near the orbifold singularity and another away from the tip so
that the vacuum expectation value becomes
\begin{eqnarray}
  \langle Z_\a \rangle = {\rm diag}(u_\a,u_\a,\cdots u_\a,v \d_{\a4})\ .
\end{eqnarray}
One can show that the mass formula again takes the form (\ref{massformula2}).
In order to study the 1/2 BPS Wilson line, we are interested in the low-energy
dynamics of those massive particles interacting with the ABJM model living on  $N-1$ M2-branes near the tip.
In the next section, it will be discussed in details.

\section{Low-energy Dynamics of Heavy Particles in M2 Theory}

More precisely, let us start with $U(N)\times U(N)$ ABJM model
and separate a single M2-brane far away from the rests sitting at the
origin of $\mathbb{C}^4/\mathbb{Z}_k$ by giving some expectation values
to complex scalar fields $Z_\a$.
For a 1/2-BPS Wilson line operator, the suitable choice
of vacuum expectation value turns out to be
\begin{eqnarray}\label{vev}
  \langle Z_{\hat \a} \rangle = 0\ , \qquad
  \langle Z_4 \rangle =\text{diag}\big(0,0,...,v\big)\ ,
  \qquad (\hat \a=1,2,3)
\end{eqnarray}
in order to preserve $SU(3) \subset SU(4)_\text{R}$.
The gauge group $U(N)\times U(N)$ is obviously
broken down to $U(N-1)\times U(N-1)$.

As will be presented in order, there are massive super-multiplets
arising from the standard Higgs mechanism at the above particular
point (\ref{vev}) on the Coulomb branch.
Those massive modes are coming from the off-diagonal modes
\begin{eqnarray}
  \Big\{ (A_\mu)_{mN},  \ (Z_{\hat \a})_{mN}, \ (\Psi^\a)_{mN} \Big\}\ ,
  \qquad \Big\{ (\tilde A_\mu)_{N m}, \ (Z_{\hat \a})_{N m},
  \ (\Psi^\a)_{Nm}\Big\}
\end{eqnarray}
together with their complex conjugates. Here $m=1,2,..,N-1$. They
transform as $({\bf N-1},{\bf 1})$, $({\bf 1},{\bf \overline{N-1}})$,
and so on under the unbroken gauge symmetry $U(N-1)\times U(N-1)$.
The mass of massive modes is given by
\begin{eqnarray}
  m = \frac{2\pi}{k} v^2\ .
\end{eqnarray}
Note that the off-diagonal modes $(Z_{4})_{mN},(Z_4)_{Nm}$ are massless,
which can be understood as Goldstone bosons eaten by massive vector bosons.

We eventually take the limit $v \to \infty$ so that
massive off-diagonal modes behave like external charged
particles in $U(N-1)\times U(N-1)$ ABJM model.
In order to study 1/2-BPS Wilson line operators,
we are interested in the Lagrangian $\hat \CL$
that governs the low-energy dynamics of such external charged particles
interacting with $U(N-1)\times U(N-1)$ ABJM fields.
One can obtain such a Lagrangian via expanding the $U(N)\times U(N)$
ABJM Lagrangian in the limit $v \to \infty$
\begin{eqnarray}
  \CL_\text{ABJM}^{U(N)} \ \to  \
  \CL_\text{ABJM}^{U(N-1)} + \hat \CL(\text{heavy modes, light modes}) +
  \CO(1/v)\ .
\end{eqnarray}
It needs however some elaborations and careful analysis
for suitable expansion, which will be presented below.

\subsection{Non-relativistic modes}

In the limit $v \to \infty$, massive modes can be treated as
non-relativistic particles, due to the fact that one
can barely create a particle/anti-particle at rest
from each massive fields such as $(Z^{\hat\a})_{mN}$.
The analysis to obtain the Lagrangian $\CL_\text{massive}$ is
therefore inevitably similar to that performed in the non-relativistic limit
of mass-deformed ABJM model \cite{Nakayama:2009cz,Lee:2009mm}. The focus is however different
as we are interested in infinite mass limit where the
spatial gradient terms become irrelevant. In the Appendix,
we give a brief review on the infinite mass limit of
massive particles with various helicity in the free field theory.

There can be many possible non-relativistic system obtained from
a single relativistic system, depending on what kinds of
particles/anti-particles we want to keep in the non-relativistic limit.
Likewise, we will end up with several different heavy particle systems
depending on our choice.
We now discuss how to choose particle or anti-particle modes for
off-diagonal massive fields in the infinite mass limit, compatible with
particular gauge choice and preserved $\CN=3$ supersymmetry or the 1/2 of
the original supersymmetry.

In our discussions below,
we choose the unitary gauge where all Goldstone bosons
$(Z_4)_{mN}, (Z_4)_{Nm}$ with their complex conjugates are turned off.
To maintain the unitary gauge, one has to demand the
following supersymmetry transformation
\begin{eqnarray}
  \d Z_4  =    \xi_{ 4\hat \a + }  \Psi^{\hat \a}_-
  -  \xi_{ 4\hat\a- } \Psi^{\hat\a}_+  \end{eqnarray}
to vanish for $(mN)$ and $(Nm)$ matrix components. Here $\pm$
represents the helicity. Inspired by the macroscopic string or
the vortex description for 1/2 BPS Wilson line [] (which will be also
presented in Section 5), let us keep only half of the supersymmetry as follow
\begin{eqnarray}\label{susyi}
  \xi_{4\hat\a +}=(\xi^{4\hat\a}_-)^*\ , \qquad
  \xi^{\hat\a\hat\b}_+ =  \e^{\hat\a\hat\b\hat\g}  \xi_{4\hat\g +}\ .
\end{eqnarray}
It therefore implies that negative helicity modes for $\Psi^{\hat \a}$
should be turned off
\begin{eqnarray}\label{choice1}
  (\Psi^{\hat \a}_-)_{mN} = 0\ , \qquad
  (\Psi^{\hat \a}_-)_{Nm} = 0\ ,
\end{eqnarray}
and their complex conjugates. Solving the free field equation for
the above off-diagonal components,
the suitable choice of non-relativistic modes turns out to be
\begin{eqnarray}
  (\Psi^{\hat \a})_{nN} = u_- \psi_{+ n}^{\hat \a}(x) e^{-imt}\ ,
  \qquad
  (\Psi^{\hat \a})_{Nn} = u_- \tilde \psi_{+n}^{\hat \a}(x) e^{+imt}\ ,
\end{eqnarray}
and similar for their complex conjugates, as shown in the Appendix. For later convenience,
we present the explicit dependence of the amplitudes on the space-time
coordinates.

The non-relativistic modes for other massive fields can be
fully determined by requiring that they are combined to
generate $\CN=3$ vector multiplets, and together by solving
their free field equations. As shown in the Appendix,
the right choice for the non-relativistic modes are therefore
given by
\begin{eqnarray}\label{choiceii}
  (\vec A)_{nN} =\sqrt{\frac{\pi}{k} } \vec E_{-} w_{+n}(x) e^{-imt}\ ,
  &&
  (Z_{\hat\a})_{nN} = \frac{1}{\sqrt{2m}} \phi_{\hat \a n}(x) e^{-imt}\ ,
  \nonumber   \\
  (\Psi^{\hat\a})_{nN} =  u_-\psi_{+ n}^{\hat\a}(x)   e^{-imt}\ ,
  &&   (\Psi^{4})_{nN} =  u_+\psi_{-n}^{4}(x)  e^{-imt}\ ,
\end{eqnarray}
and
\begin{eqnarray}\label{choiceiii}
  (\vec{\tilde A})_{Nn} =\sqrt{\frac{\pi}{k}} \vec E_{-} \tilde w_{+n}(x)
  e^{+imt}\ ,
  &&
  (Z_{\hat\a})_{Nn} = \frac{1}{\sqrt{2m}} \tilde \phi_{\hat \a n}(x) e^{+imt}\ ,
  \nonumber   \\
  (\Psi^{\hat\a})_{Nn} =  u_- \tilde \psi_{+ n}^{\hat\a}(x)   e^{+imt}\ ,
  &&   (\Psi^{4})_{Nn} =  u_+ \tilde \psi_{-n}^{4}(x)  e^{+imt}\ ,
  \label{choice2}
\end{eqnarray}
where $\vec E_\pm = \big( 1 , \pm i \big)$ denote the polarization vectors
with definite helicity $\pm 1$. Here all normalization factors are determined
by canonical kinetic terms for heavy particles.

As promised, the non-relativistic modes $w_+, \psi^{\hat \a}_+ ,\phi_{\hat\a}$
and $\psi^4_-$ which transform as $({\bf N-1},{\bf 1})$
under the unbroken gauge symmetry $U(N-1)\times U(N-1)$ are combined to generate
the $\CN=3$ vector multiplet
\begin{eqnarray}
  \begin{array}{r|cccc}
  & w_+ & \psi^{\hat \a}_+ & \phi_{\hat \a} & \psi^4_- \\
  \hline \hline
  \text{helicity} & + 1 & +1/2 & 0 & -1/2\\
  \hline
  \text{degeneracy} & 1 & 3 & 3 & 1
  \end{array}\ \  .
\end{eqnarray}
Similarly, the non-relativistic modes which furnish $({\bf 1},{\bf N-1})$
representation are combined to generate another $\CN=3$ vector multiplet
\begin{eqnarray}
  \begin{array}{r|cccc}
  & \bar{\tilde w}_- & \bar{\tilde \psi}_{\hat \a- } &
  \bar{\tilde \phi}^{\hat \a} & \bar{\tilde \psi}_{4+} \\
  \hline \hline
  \text{helicity} & - 1 & -1/2 & 0 & +1/2\\
  \hline
  \text{degeneracy} & 1 & 3 & 3 & 1
  \end{array}\ \  .
\end{eqnarray}

\subsection{Low-energy dynamics of 1/2-BPS particles}

We present in this section the detailed steps to
obtain the low-energy Lagrangian $\hat \CL$ for the
non-relativistic modes. Basically all we need to do is
to insert (\ref{choiceii}, \ref{choiceiii})  into the
$U(N)\times U(N)$ ABJM model and expand it to read
off the leading terms. The non-trivial interactions between
heavy 1/2-BPS particles with $U(N-1)\times U(N-1)$ ABJM model
however leads to several delicate points we should take into account.
In particular some $1/v$-corrections to the
non-relativistic modes (\ref{choiceii}, \ref{choiceiii}),
obtained in the free field theory limit, can arise.

We hereafter use abusing notations for massless $U(N-1)\times U(N-1)$ ABJM fields.
They will be denoted by $Z_\a, \Psi^\a, A_\mu$ and $\tilde A_\mu$
just like original $U(N)\times U(N)$  ABJM fields unless any confusion arises.

\paragraph{Scalar parts} Let us first blindly expand the bosonic potential
to quadratic order in the off-diagonal massive components which could
survive in the infinite mass limit
\begin{eqnarray}\label{potential}
 U^{U(N)}  \!\!\! &=& \!\!\!  U^{U(N-1) }  + m^2
  \bar \Phi^{\hat\a} \Big( 1+ \frac{2}{v^2}  \O^{\b}_{\ \g}  Z_\b \bar Z^\g  \Big)
  \Phi_{\hat\a} + m^2 \tilde \Phi^{\hat\a} \Big( 1+\frac{2}{v^2}  \O^{\g}_{\ \b}
  \bar Z^\b   Z_\g \Big) \bar{\tilde\Phi}_{\hat\a}     \nonumber \\
  & &   \!\!\! + \frac{m^2}{v}
  \Big( \tilde \Phi_4 \bar Z^{\hat\a} \Phi_{\hat\a} +
  \bar\Phi^{\hat\a} Z_{\hat\a} \bar{\tilde\Phi}^4 +
  \tilde \Phi_{\hat\a} \bar Z^{\hat\a} \Phi_4  +
  \bar \Phi^4 Z_{\hat\a} \bar{\tilde\Phi}^{\hat\a} \Big) \nonumber \\
  & & \!\!\!-\frac{m^2}{v^2}  \Big(  \bar\Phi^{\hat\a} Z_{\hat\a}
  \bar Z^{\hat\b} \Phi_{\hat\b}  -\bar\Phi^4 Z_{\hat\a} \bar Z^{\hat\a} \Phi_4 +
  \tilde\Phi_{\hat\a} \bar Z^{\hat\a} Z_{\hat\b} \bar{\tilde \Phi}^{\hat\b}
  -\tilde\Phi_4 \bar Z^{\hat\a} Z_{\hat\a} \bar{\tilde\Phi}^4 \Big)\ ,
\end{eqnarray}
where $\O^\a_{\ \b} = \text{diag}(1,1,1,-1)$ and $[\Phi_\a]_n=(Z_\a)_{nN}$,
$[\bar{\tilde\Phi}^\a]_n = (\bar Z^\a)_{nN}$. The unitary gauge is not
imposed yet for a reason clarified below.
Non-trivial interactions in the last two lines of (\ref{potential})
lead to subleading corrections to the mass-eigenstates for scalar fields.
It implies that the non-relativistic modes (\ref{choiceii}, \ref{choiceiii})
get slightly modified by
\begin{eqnarray}
 \Big( \delta_{\hat\a}^{\hat\b} - \frac{1}{2v^2} Z_{\hat\a}\bar Z^{\hat\b}\Big)
 \Phi_{\hat\b} + \frac{1}{v} Z_{\hat\a} \bar{\tilde\Phi}^4&=&
 \frac{1}{\sqrt{2m}} \phi_{\hat\a} e^{-imt} \ , \nonumber \\
 \Big( \delta^{\hat\a}_{\hat\b} - \frac{1}{2v^2} \bar Z^{\hat\a} Z_{\hat\b}\Big)
 \bar{\tilde\Phi}^{\hat\b} + \frac{1}{v} \bar Z^{\hat\a}  \Phi^4 &=&
 \frac{1}{\sqrt{2m}} \bar{\tilde \phi}_{\hat\a} e^{-imt}\ ,
\end{eqnarray}
and the unitary gauge $[Z_4]_{nN}= [Z_4]_{Nn}=0$ is also rotated by
\begin{eqnarray}
 0 = [\CZ_4]_{nN} &\equiv&
 \Big( 1  - \frac{1}{2v^2} Z_{\hat\a}\bar Z^{\hat\a}\Big) \Phi_4
 -\frac{1}{v} Z_{\hat\a} \bar{\tilde\Phi}^{\hat\a}\ , \nonumber \\
 0 = [\bar \CZ_4]_{nN} &\equiv&
 \Big( 1  - \frac{1}{2v^2} \bar Z^{\hat\a}  Z_{\hat\a}\Big)
 \bar{\tilde \Phi}_4 - \frac{1}{v} \bar Z^{\hat\a}  \Phi_{\hat\a}\ .
\end{eqnarray}
The above redefinition of the scalar fields can be understood as
an infinitesimal $SU(4)_\text{R}$ rotation with the parameter $\bar Z^{\hat a}/v$.
We therefore end up with canonical kinetic terms and interactions
for external scalar particles
\begin{eqnarray}\label{scalar}
  \hat \CL_\text{scalar} = i \bar \varphi^{\hat \a} D_0 \varphi_{\hat \a}
  + i \tilde \varphi_{\hat \a} D_0 \bar{\tilde \varphi}^{\hat \a}
  - \frac{2\pi}{k} \Big[ \bar \varphi^{\hat \a}
  \Big( \O^{\b}_{\ \g}  Z_\b \bar Z^\g \Big) \varphi_{\hat \a}
  + \tilde \varphi_{\hat \a}
  \Big( \O_{\ \b}^{\g}  \bar Z^\b Z_\g \Big) \bar{ \tilde \varphi}^{\hat \a} \Big]\ .
\end{eqnarray}
Note that the interactions between massless ABJM scalars and heavy scalars
are in a perfect matching with the mass formula (\ref{massformula2}).

\paragraph{Vector parts} Expanding the scalar kinetic terms
gives us the interactions between W-bosons and massless scalar fields
\begin{eqnarray}
  \CL_\text{vector}^\text{int} = - \big| \bar Z^{\hat \a} \vec W \big|^2
  - \big| Z_{\hat \a}\vec{\tilde W}\big|^2
  - v^2 \Big| \vec{W} - \frac{Z_4}{v} \vec{\tilde W} \Big|^2
   -v^2 \Big| \vec{\tilde W} - \frac{\bar Z^4}{v}  \vec{W}\Big|^2 +
   \CO(1/v) \ ,
\end{eqnarray}
where $(\vec A)_{mN} = \vec W_m$ and $(\vec{\tilde A})_{mN}= \vec{\tilde W}_m$.
They imply again that mass-eigenstates for vector bosons (\ref{choiceii},
\ref{choiceiii}) also get shifted by
\begin{eqnarray}
  \Big( 1 + \frac{Z_4\bar Z^4}{2v^2} \Big) \vec W - \frac{Z_4}{v} \vec{\tilde W}
  &=&  \sqrt{\frac{\pi}{k}} \vec E_{-} w_+ e^{-imt}\ ,
  \nonumber \\
  \Big( 1 + \frac{\bar Z_4 Z^4}{2 v^2} \Big) \vec{\tilde W} - \frac{\bar Z^4}{v}  \vec{W}
  &=&  \sqrt{\frac{\pi}{k}}\vec E_{+} \bar{\tilde w}_-  e^{-imt}\ .
\end{eqnarray}
This transformation is indeed an infinitesimal Lorentz transformation
on the field space $(\vec W, \vec{\tilde W})$, leaving the Chern-Simons
kinetic terms intact
\begin{eqnarray}
  \CL^{U(N)}_\text{CS} &=& \CL^{U(N-1)}_\text{CS}
  + i \bar w_- D_0 w_+   + i \tilde w_+ D_0 \bar{\tilde w}_-
  + m \big( \bar  w_-  w_+ +
  \tilde  w_+ \bar{\tilde  w}_- \big)\ .
\end{eqnarray}
The Lagrangian for massive vector boson can therefore takes the following form
\begin{eqnarray}\label{vector}
  \hat \CL_\text{vector} = i \bar w_- D_0 w_+
  + i \tilde w_+ D_0 \bar{\tilde w}_-
  - \frac{2\pi}{k} \Big[\bar w_- \big( \O^\a_{\ \b} Z_{\a}
  \bar Z^{\b} \big) w_+ + \tilde w_+ \big( \O^\b_{\ \a}
  \bar Z^{\a} Z_{\b} \big)
  \bar{\tilde w}_- \Big] .
\end{eqnarray}

\paragraph{Fermion parts}

In order to find out proper mass-eigenstates for fermion
fields, it needs much careful analysis to expand the Yukawa
interactions, which turns out to be little tricky. One can
show that non-trivial interactions again lead to
the sub-leading corrections to the mass-eigenstates as below
\begin{eqnarray}
  \Big[ \Big( \d^{\hat\a}_{\hat\b} -\frac{1}{2v^2}
  \big(  Z_{\hat\g}\bar Z^{\hat\g}\d^{\hat\a}_{\hat\b} -
  Z_{\hat\b} \bar Z^{\hat\a} \big) \Big) \Psi^{\hat\b} +
  \frac{1}{v} \e^{\hat\a\hat\b\hat\g} Z_{\hat\b}\bar\Psi_{\hat\g}\Big]_{nN}
  &=& u_- \psi^{\hat\a}_{+ n}(x) e^{-imt}\ , \\
 \Big[\Big( \d^{\hat\b}_{\hat\a} -\frac{1}{2v^2}
 \big(  \bar Z^{\hat\g} Z_{\hat\g}\d^{\hat\b}_{\hat\a} -
 \bar Z^{\hat\b} Z_{\hat\a}\big) \Big) \bar\Psi_{\hat\b} +
 \frac{1}{v} \e_{\hat\a\hat\b\hat\g} \bar Z^{\hat\b} \Psi^{\hat\g}\Big]_{nN}
 &=& u_+ \bar{\tilde\psi}_{\hat\a - n}(x) e^{-imt}\ .  \nonumber
\end{eqnarray}
In terms of modified mass-eigenstates for fermions,
the Yukawa coupling can be expanded as
\begin{eqnarray}
  \CL_\text{Yukawa}^{U(N)} &=& \CL_\text{Yukawa}^{U(N-1)}
  - m(\bar\psi_\a\psi^\a + \tilde\psi^\a\bar{\tilde\psi}_\a )
  - \frac{2\pi}{k} \Big[ \bar \psi_{\a}
  \big(\O^\a_{\ \b} Z_{\a} \bar Z^{\b} \big) \psi^{\a}
  + \tilde \psi^{\a} \big( \O^\b_{\ \a} \bar Z^{\a} Z_{\b} \big)
  \bar{\tilde \psi}_{\a}\Big] \nonumber \\
  &&  \!\!\!\!  + \sqrt{\frac{4\pi}{k}}\Big[
  \bar \phi^{\hat\a} \Psi^4_+ \bar{\tilde\psi}_{\hat\a-}  +
  \bar\psi_{\hat\a-} \Psi^4_+ \bar{\tilde\phi}^{\hat\a}   +
  \tilde\phi_{\hat\a} \bar\Psi_{4-} \psi^{\hat\a}_+  +
  \tilde\psi^{\hat\a}_+ \bar\Psi_{4-} \phi_{\hat\a}  \Big] + \CO(\frac{1}{\sqrt{m}} )\ .
\end{eqnarray}
For clarity, we hereafter sometimes drop the helicity indices unless
it does not make any confusion. The fermion kinetic terms can be
expanded as below
\begin{eqnarray}
  \CL_\text{f. kin}^{U(N)} &=& \CL_\text{f. kin}^{U(N-1)} + m \big(
  \bar \psi_{\a}  \psi^{ \a}   +
  \tilde \psi^{\a} \bar{\tilde \psi}_{\a} \big)
  + i\bar\psi_{\a} D_0 \psi^{\a}
  + i\tilde{\psi}^{\a } D_0 \bar{\tilde \psi}_{\a}
  \nonumber \\
  &&
  + \sqrt{\frac{4\pi}{k}} \Big[  \tilde \w_+ \bar\Psi_{4-} \psi^{4}_-
  + \tilde\psi^4_- \bar\Psi_{4-} \w_+
  + \bar\psi_{4+} \Psi^4_+ \bar{\tilde \w}_-
  + \bar \w_- \Psi^4_+ \bar{\tilde\psi}_{4+}\Big] + \CO(\frac{1}{\sqrt{m}})\ . \nonumber
\end{eqnarray}
As a consequence, the Lagrangian for heavy fermions becomes
\begin{eqnarray}\label{fermion}
  \hat \CL_\text{fermion} &=& i\bar\psi_{\a} D_0 \psi^{\a}
  + i\tilde{\psi}^{\a } D_0 \bar{\tilde \psi}_{\a}
  - \frac{2\pi}{k} \Big[ \bar \psi_{\a}
  \big(\O^\a_{\ \b} Z_{\a} \bar Z^{\b} \big) \psi^{\a}
  + \tilde \psi^{\a} \big( \O^\b_{\ \a} \bar Z^{\a} Z_{\b} \big)
  \bar{\tilde \psi}_{\a}\Big]
  \nonumber \\
  &&
  + \sqrt{\frac{4\pi}{k}} \Big[  \tilde \w_+ \bar\Psi_{4-} \psi^{4}_{-}
  + \tilde\psi^4_- \bar\Psi_{4-} \w_+
  + \bar\psi_{4+} \Psi^4_+ \bar{\tilde \w}_-
  + \bar \w_- \Psi^4_+ \bar{\tilde\psi}_{4+}\Big]
  \nonumber \\
  &&
  + \sqrt{\frac{4\pi}{k}}\Big[
  \bar \phi^{\hat\a} \Psi^4_+ \bar{\tilde\psi}_{\hat\a-}  +
  \bar\psi_{\hat\a-} \Psi^4_+ \bar{\tilde\phi}^{\hat\a}   +
  \tilde\phi_{\hat\a} \bar\Psi_{4-} \psi^{\hat\a}_+  +
  \tilde\psi^{\hat\a}_+ \bar\Psi_{4-} \phi_{\hat\a}  \Big]\ .
\end{eqnarray}

\paragraph{Summary} Collecting the results
(\ref{scalar}), (\ref{vector}) and (\ref{fermion}),
the low-energy dynamics of external 1/2-BPS particles
interacting with ABJM fields are governed by
\begin{eqnarray}\label{fulllag}
  \hat \CL &=&
  i \bar \w_- \CD_0 \w_+  + i \tilde \w_+ \tilde D_0 \bar{\tilde \w}_-
  + i\bar\psi_{\a} \CD_0 \psi^{\a}
  + i\tilde{\psi}^{\a } \tilde \CD_0 \bar{\tilde \psi}_{\a}
  + i \bar \varphi^{\hat \a} \CD_0 \varphi_{\hat \a}
  + i \tilde \varphi_{\hat \a} \tilde \CD_0 \bar{\tilde \varphi}^{\hat \a}
  \nonumber \\
  &&
  + \sqrt{\frac{4\pi}{k}} \Big[  \tilde \w_+ \bar\Psi_{4-} \psi^{4}_{-}
  + \tilde\psi^4_- \bar\Psi_{4-} \w_+
  + \bar\psi_{4+} \Psi^4_+ \bar{\tilde \w}_-
  + \bar \w_- \Psi^4_+ \bar{\tilde\psi}_{4+}\Big]
  \nonumber \\
  &&
  + \sqrt{\frac{4\pi}{k}}\Big[
  \bar \phi^{\hat\a} \Psi^4_+ \bar{\tilde\psi}_{\hat\a-}  +
  \bar\psi_{\hat\a-} \Psi^4_+ \bar{\tilde\phi}^{\hat\a}   +
  \tilde\phi_{\hat\a} \bar\Psi_{4-} \psi^{\hat\a}_+  +
  \tilde\psi^{\hat\a}_+ \bar\Psi_{4-} \phi_{\hat\a}  \Big]\ ,
\end{eqnarray}
where the covariant derivatives are defined as
\begin{eqnarray}
  \CD_0 = \partial_0 - i \CA_0 \ , \qquad
  \CA_0 = A_0 - \frac{2\pi}{k} \O^\a_{\ \b} Z_{\a} \bar Z^{\b} \ ,
  \nonumber \\
  \tilde \CD_0 = \partial_0 - i \tilde \CA_0 \ , \qquad
  \tilde \CA_0 = \tilde A_0 -\frac{2\pi}{k}  \O^\b_{\ \a} \bar Z^{\a} Z_{\b}\ .
\end{eqnarray}

\section{Half BPS Wilson Line in M2-Theory}

It is now ready to discuss the half BPS Wilson line
with focus on physical origin of the superconnection.
Let us begin by managing the low-energy Lagrangian for
heavy particles (\ref{fulllag}) into an appealing
expression
\begin{eqnarray}\label{fulllagi}
  \hat \CL = \text{Tr}\Big[ i
  \bar \varPsi^\a \hat D_0 \varPsi_\a \Big]\ ,
\end{eqnarray}
where $\varPsi_\a$ are supermatrices defined as
\begin{eqnarray}
  \varPsi^{\hat \a} =
  \begin{pmatrix}
    \varphi_{\hat \a} & \psi^{\hat \a}_+ \\
    \bar{\tilde \psi}_{\hat \a -} & \bar{\tilde\varphi}^{\hat \a}
  \end{pmatrix}\ ,
  \qquad
  \varPsi^{4} =
  \begin{pmatrix}
    \w_+ & \psi^{4}_- \\
    \bar{\tilde \psi}_{4 +} & \bar{\tilde \w}_-
  \end{pmatrix}\ ,
\end{eqnarray}
and $\hat D_0$ represent a super-covariant derivative with
an $U(N|N)$ superconnection $\hat A_0$
\begin{eqnarray}
  \hat D_0 = \partial_0 - i \hat A_0\ ,
  \qquad
  \hat A_0 = \begin{pmatrix}
    \CA_0 & \sqrt{\frac{4\pi}{k}} \Psi_+^4 \\
    \sqrt{\frac{4\pi}{k}} \bar \Psi_{4-} & \tilde \CA_0
  \end{pmatrix}\ .
\end{eqnarray}
It implies that the low-energy dynamics of 1/2-BPS massive particles
respects the supergauge symmetry $U(N|N)$, provided that the matter fields
$\varPsi_\a$ and superconnection $\hat A_0$ transforms as
\begin{eqnarray}
  \varPsi  \ \to \ U^\dagger \varPsi_\a\ , \ \
  \hat A_0 \ \to \ U^\dagger \hat A_0 U + i U^\dagger \partial_0 U \ ,
  \qquad
  U = e^{- i \L} \in U(N|N)\ .
\end{eqnarray}

As mentioned repeatedly, one can understand the 1/2-BPS Wilson line  evolve under the interactions to ABJM model. Since the equations of
motions for massive particles are
\begin{eqnarray}
  \hat D_0 \varPsi_\a  = 0\ ,
\end{eqnarray}
the time-evolution factor of the wavefunctions, or 1/2-BPS Wilson
line is given by
\begin{eqnarray}
  \text{W}(t) = \CP \text{exp} \Big[ i \int^t d\t \
  \hat A_0 \Big] \ ,
\end{eqnarray}
which exactly matches with the result of \cite{Drukker:2009hy}.
The supergauge symmetry of (\ref{fulllagi}) explains
the physical origin of the form of the 1/2-BPS Wilson line
as the holonomy of $U(N|N)$ superconnection.
Free constant spinor parameters $\eta_\a$  of \cite{Drukker:2009hy} can be also
understood as helicity projections of massive 1/2-BPS
particles in the infinite mass limit.

Note that the wavefunctions evolves by mixing the particles
in different $\CN=3$ vector multiplets. It strongly
implies that the Wilson line could be invariant under
the supersymmetry parameters complement to (\ref{susyi})
\begin{eqnarray}\label{susyii}
  \xi^{\hat \a 4}_+ \ , \qquad
  \xi^{\hat \a \hat \b}_-\ ,
\end{eqnarray}
which also mixes particles in different $\CN=3$ vector
multiplets. It indeed turns out to be the case. Under the
supersymmetry transformation with (\ref{susyii}),
one can show that the superconnection $\hat A_0$ transforms
as
\begin{eqnarray}
 \d_\text{SUSY} \hat A_0 = \partial_0 \L - i \big[ \hat A_0, \L \big]
 \label{susygauge}
\end{eqnarray}
with
\begin{eqnarray}
  \L = \sqrt{\frac{4\pi}{k}} \begin{pmatrix}
    0 &  iZ_{\hat \a} \xi^{\hat \a 4}_+ \\
    -i\bar Z^{\hat \a} \xi_{\hat \a 4- } & 0
  \end{pmatrix}\ .
\end{eqnarray}
It is nothing but a specific supergauge transformation
with parameter $\L \in u(N|N)$. The Wilson line operator
\begin{eqnarray}
  \text{W}(t_f,t_i) = \CP \text{exp} \big[ i \int^{t_f}_{- t_i}  d\t \
  \hat A_0 \big]\ ,
\end{eqnarray}
would transform covariantly under the supersymmetry
\begin{eqnarray}
  {\rm W}(t_f,t_i)  \rightarrow U(t_f)^\dagger {\rm W} (t_f,t_i) U(t_i) \ .
\end{eqnarray}
For a closed loop, there are two possible periodic boundary conditions on
the fermion fields $\Psi^4_+$ in the superconnection $\hat A$.
The supersymmetric transformation (\ref{susygauge}) in turn decides
the boundary condition on $\xi^{\hat\a 4}_+$.
For the periodic boundary condition, the 1/2-BPS Wilson loop involves supertrace,
\begin{eqnarray}
  {\rm W}_\text{periodic} = {\rm STr} \
  \CP \text{exp} \big[ \oint d\t \
  \hat A_0 \big]
\end{eqnarray}
while, for the anti-periodic one, taking the ordinary trace gives us the
1/2-BPS Wilson loop operator
\begin{eqnarray}
  {\rm W}_\text{anti-periodic} = {\rm Tr} \ \CP \text{exp} \big[ \oint d\t \
  \hat A_0 \big] \ .
\end{eqnarray}
In \cite{Drukker:2009hy}, it has been argued that the proper boundary condition
for the circular Wilson loop in Euclidean $\IR^3$ is the anti-periodic boundary
condition, i.e., the ordinary trace leads to the supersymmetric Wilson loop.

\section{1/2-BPS Vortices and External Particles}

In the previous section we showed how the Wilson line operator arises
when the external particles interact with the ABJM model.
Let us now in turn think of a different aspect of the Wilson line operator.
In particular, we are interested in how some
classical pictures can be affected when the Wilson line
operator is introduced in the path integral formulations
of low-energy theory on M2-branes. Similar to our analysis below
has been studied in \cite{Drukker:2008jm} for different purposes.

Let us start with the $U(1)\times U(1)$ ABJM model
whose bosonic Lagrangian takes the following form
\begin{eqnarray}
  {\cal L}^{(1)}_\text{abelian} &=&  -D_\mu \bar{Z}^\a D^\mu Z_\a
  + \frac{k} {2\pi} \epsilon^{\mu\nu\rho}  b_\mu \partial_\nu c_\rho \nonumber  \\
  &=&  - \big|(\partial_\mu-ib_\mu )Z_\a\big|^2 +
  \frac{1}{4\pi} \epsilon^{\mu\nu\rho} ( k b_\mu - \partial_\mu\sigma) f_{\nu\rho}\ ,
  \label{abelian1}
\end{eqnarray}
where $b_\mu = A_\mu - \tilde{A}_\mu$ and $c_\mu = (A_\mu+ \tilde{A}_\mu)/2 $.
For the last equality, one introduces an auxiliary two-form field $f_{\mu\nu}$.
The invisibility of magnetic monopoles of $2\pi$ $f_{12}$-flux demands
the scalar field $\s$ to have $2\pi$ periodicity $\sigma \sim \sigma+2\pi$.
Integrating over $f_{\mu\nu}$, one can rewrite (\ref{abelian1})
into the almost free Lagrangian
\begin{eqnarray}
  {\cal L}^{(1)}_\text{abelian} = - \big| \partial_\mu \hat Z_\a \big|^2 \ ,
  \qquad \hat Z_\a = e^{i\s/k} Z_\a
\end{eqnarray}
with the residual gauge symmetry $\hat Z_\a \sim e^{2\pi i n /k}\hat Z_\a$,
which leads to the moduli space of a single  M2 brane
\begin{eqnarray}
  \CM = {\mathbb{C}^4}/\mathbb{Z}_k \ .
\end{eqnarray}

\paragraph{1/2-BPS vortex}

We first study a classical 1/2-BPS bosonic object
in the $U(1)\times U(1)$ ABJM model for simplicity.
Looking at the SUSY variation rules for fermions
\begin{eqnarray}
  \d\Psi_+^\a &=& -i   \xi^{\a\b}_+ D_0Z_\b
  +i \xi^{\a\b}_- D_+Z_\b \ ,
  \nonumber \\
  \d \Psi^\a_-&=& +i\xi^{\a\b}_-D_0 Z_\b -i \xi^{\a\b}_+ D_- Z_\b \ ,
  \end{eqnarray}
one can show that vortex solutions satisfying the following
equations
\begin{eqnarray}
  Z_1=Z_2=Z_3=0\ , \  \
  D_0 Z_4=0\ , \ \ D_+ Z_4 =0\ , \ \ D_- Z_4 \neq 0\ ,
  \label{12bps}
\end{eqnarray}
preserves half of the supersymmetry along $\xi^{\hat\a 4}_-, \xi^{\hat\a\hat\b}_+$.
Here $D_\pm = D_1\pm i D_2$ and $A_\pm = A_1\pm i A_2$.
The field equations for gauge bosons are given by
\begin{eqnarray}
 \frac{k}{2\pi} F_{i0} + D_i (Z_4 \bar  Z^4) = 0\ , &&  F_{12}= 0 \ , \nonumber  \\
 \frac{k}{2\pi} \tilde F_{i0} + D_i (\bar Z^4 Z_4 ) = 0 \ ,  &&
 \tilde  F_{12}=0\ .
\end{eqnarray}
For static solutions, they can be solved by
\begin{eqnarray}
  A_0=\tilde A_0 = - \frac{2\pi}{k} |Z_4|^2\ .
\end{eqnarray}

In terms of the gauge invariant variable $\hat Z_4$,
the 1/2-BPS vortex equation $\partial_+ \hat Z_4=0$
implies that the vortex configuration can be
described as a holomorphic function.
Due to the residual gauge symmetry $\IZ_k$ ($\hat{Z}_4\sim e^{2\pi i/k} \hat{Z}_4$),
the 1/2-BPS elementary vortex becomes
\begin{eqnarray}\label{evortex}
  \hat Z_4 = \frac{p_v}{(z-z_0)^{1/k}} \ ,
\end{eqnarray}
where $z_0$ denotes the position of the source.
Note that the dimensionful parameter $p_v$ indicates the
complicated internal structure of the solution\footnote{
There is also a 1/2-BPS funnel solution $ \hat Z_4 = c_f z^{1/k}$
which has different boundary condition.}.

The elementary vortex describes how the world-sheet of single M2-brane
is deformed by the external point source. The configuration
(\ref{evortex}) implies that the M2-brane is pulled by the
source to the spatial infinity, and wraps the $S^1$ fibre of
the orbifold $\IC^4/Z_k$ once. The source can be therefore identified as
an infinitely long type IIA fundamental string,
qualitatively similar to heavy particles of the Wilson line.

This elementary vortex has infinite energy due to its
singularity at the point $z_0$. Although the gauge fields
should be pure gauge, they can carry the non-zero
flux at the source: suppose that
\begin{eqnarray}
  A_- =\frac{-i \a}{z-z_0}\ , \qquad
  \tilde A_- = \frac{-i \b}{z-z_0} \ ,
\end{eqnarray}
which describe point magnetic fluxes $2\pi \a$ and $2\pi \b$ at the source $z_0$.
The BPS equations for vortex solutions becomes
\begin{eqnarray}
  D_+ Z_4= (\p_+ + \frac{\a-\b}{\bar z -\bar z_0} )Z_4 = 0\ ,
  \qquad
  D_- Z_4= (\p_- + \frac{\a-\b}{z - z_0} )Z_4 = 0\ .
\end{eqnarray}
In the abelian Higgs system, the vorticity and magnetic flux are
tightly correlated when we require the finite energy configurations
so that the gauge invariant scalar has net-zero vorticity.
Our vortex is somewhat different as the scalar $Z_4$ vanishes at the spatial infinity.
All we need to require is that the gauge invariant $\hat Z_4$ carries $1/k$ vorticity.
The natural choice for the 1/2-BPS external source then generates
two types of vortices as
\begin{eqnarray}
  Z_4= \frac{c}{|z-z_0|^{1/k}}\ , && A_- = - \frac{i}{k(z-z_0)}  \ \ \
  (A-{\rm type \ vortex})\ , \nonumber  \\
  Z_4= \frac{c}{|z-z_0|^{1/k}}\ , && \tilde A_- = + \frac{i}{k(z-z_0)}
  \  \ \ (\tilde A-{\rm type\ vortex})\ .
\end{eqnarray}

\paragraph{Dual description of ABJM model}

In order to study another aspect of the 1/2-BPS vortex,
let us present a dual description of the ABJM model
with use of the vector-scalar duality, so-called Mukhi-Papageorgakis
map \cite{Mukhi:2008ux}.

Integrating over the gauge field $b_\mu$ in (\ref{abelian1}),
one can obtain
\begin{eqnarray}\label{aaa}
  C_{\mu\nu} \equiv  \p_\mu c_\n -\p_\n c_\m  =
  2 \e_{\m\n\r}\sum_\a  \big| Z_\a\big|^2\big( \p^\r \text{arg}Z_\a - b^\r
  \big) \ .
\end{eqnarray}
Since we are now interested in the case where all scalars
except $Z_4$ are turned off, $b_\mu$ can be expressed by
\begin{eqnarray}
  b_\mu = \frac{1}{\phi}\Big[
  \partial_\m \text{arg}Z_4 - \frac14 \e_{\mu\nu\rho} C^{\n\r}\Big]\ ,
  \qquad \phi= \frac{2\pi}{k} |Z_4|^2\ .
\end{eqnarray}
In terms of $C_{\mu\nu}$ and $\phi$, the Lagrangian (\ref{abelian1}) can be
rewritten as
\begin{eqnarray}\label{abelian2}
  {\cal L}_\text{abelian}^{(2)} =
  -\frac{k}{8\pi\phi} (\p_\m \phi)^2 - \frac{k}{16\pi \phi} C_{\mu\nu}C^{\mu\nu}\ .
\end{eqnarray}
It is noteworthy here that $\phi$ and $c_\mu$ are in fact combined to
form $\CN=2$ vector multiplet in three dimensions. For more
systematic analysis in a manifestly supersymmetric fashion, it
is referred to \cite{Koh:2009um}.

\paragraph{1/2-BPS vortex revisited}
In terms of the $\phi$ and $c_{\mu}$   variables,
the 1/2 BPS vortices can be characterized by the solution
\begin{eqnarray}\label{vortex}
  \phi = -c_0  = \frac{2\pi}{k}\frac{p_v^2}{ |z|^{2/k}}
\end{eqnarray}
with fluxes at the source. One can roughly understand that
it is the BPS solution from the complete squares of energy density
\begin{eqnarray}
 \CE = \frac{k}{8\pi \phi}
 \Big((\p_i \phi)^2 + C_{i0}^2\Big) = \frac{k}{8\pi \phi}
 (\p_i \phi + C_{i0})^2 - \p_i \left( \frac{k}{4\pi} C_{i0} \right) \ ,
\end{eqnarray}
up to the Gauss law. Here we ignored the source term.
In order to generate the 1/2-BPS vortex (\ref{vortex}),
one should add to the Lagrangian (\ref{abelian2}) the point source terms as
\begin{eqnarray}
  \CL_\text{source}=  (\phi+ c_0)\d^2(z) + \cdots \ ,
  \label{source}
\end{eqnarray}
where we do not specify the source terms for the flux yet.

Let us consider the insertion of the $A$-type 1/2-BPS Wilson-line to the abelian theory,
ignoring fermionic contributions. It introduce the source terms below
\begin{eqnarray}\label{source1}
  \CL^{A-{\rm type}}_\text{source} &=& \big( A_0 -\frac{2\pi}{k}
  ( |Z_{\hat\a}|^2 -|Z_4|^2) \big)\d^2(z)  \nonumber   \\
  &=& (c_0 + \phi)\d^2 (z) + \frac12 b_0  \d^2 (z)
  -\frac{2\pi}{k}  |Z_{\hat\a}|^2  \d^2(z)
\end{eqnarray}
Both the scalar field equation and the energetic consideration imply that
$Z_{\hat\a}=0$ to prevent the energy increase. Here $A_0=c_0 + b_0/2$.
One can also show that the second terms of (\ref{source1}) is necessary
to generate the $f_{12}$ flux at the source $f_{12}= \pi \d^2(x)/k$.
The above source terms can therefore be identified as the source
terms (\ref{source}) for 1/2-BPS vortices.
This analysis confirms that the insertion of 1/2 BPS Wilson line parallels to
the insertion of 1/2 BPS vortices.

The multi-vortex solutions are given by
\begin{eqnarray}
\hat{Z}_4 \equiv f(z)=  \frac{p_v}{\prod_p(z-z_p)^{1/k} }\ .
\end{eqnarray}
where the parameter $p_v$ is dimensionful in general.
Note that the multi-vortex solution is multiplicative, not additive.
When two vortices overlap, the factor therefore becomes $1/z^{2/k}$ so that
the winding over the $S^1$ fibre is doubled.
In the large k limit, one can show both vortex and funnel solutions
become the logarithmic solutions similar to the reaction of
a D2-brane under the external source.

Only for $k=2$, a single vortex solution can be scale invariant due to the fact that
the coefficient $p_v$ become dimensionless.
For even $k$, $k/2$ vortices on the top of each other is scale invariant.
Further properties of the scale invariant solution has been investigated in \cite{Drukker:2008jm}.

Ultimately we are interested in the insertion of non-abelian Wilson line
operators with their quantum nature. It would be very interesting to see how
such structure parameters would survive and manifest.
At least from the string calculation of particle-antiparticle attractive force
in finite distance, done in \cite{Drukker:2008zx}, the attractive energy is given by
\begin{eqnarray}
  E\sim  -\frac{1}{L}\sqrt{\frac{N}{k}}
\end{eqnarray}
which is blind to the structures of vortices. It is also falling off faster than
the fall off $1/L^{2/k}$, naively expected for vortex/anti-vortex attraction.

\vskip 1cm
\centerline{\bf Acknowledgement}
\vskip 5mm
\noindent
We would like to thank Nakwoo Kim, Sangmin Lee, and Piljin Yi
for valuable discussions. We also thank Jun-Bao Wu for collaboration
at an early stage of this work. This work (KML) is supported in part
by the Center for Quantum Spacetime of Sogang University with grant number
{\bf R11-2005-021} and the Korea National Research Foundation
({\bf No}.2006-0093850, {\bf No}. 2009-0084601).

\newpage

\centerline{\Large \bf Appendix}

\appendix

\section{Infinite Mass Limit in Free Field Theories}

The infinite mass limit $v\to \infty$ somehow parallels with the
standard non-relativistic limit $c\to \infty$, due to
the fact that factors $(\vec p/mc)$ and $(E/mc^2)$ become negligible
in both limits. Based on the analysis for the non-relativistic
limit of free massive particles \cite{Nakayama:2009cz,Lee:2009mm}, we present how to
take the infinite mass limit for those particles.

\subsection{Scalar field}

Let us begin by a Lagrangian for a free massive scalar
\begin{eqnarray}
  \CL^\text{scalar} =  D_0 \bar{Z} D_0  Z  - D_i \bar{Z} D_i  Z
  - m^2 \bar{Z} Z\ .
\end{eqnarray}
Considering a particle mode in the scalar field $Z$
\begin{eqnarray}
  Z = \frac{1}{\sqrt{2m}} \ \phi (t,{\bf x}) e^{-imt}\ ,
\end{eqnarray}
the above Lagrangian in the  limit $m\to \infty$
becomes
\begin{eqnarray}
  \CL^\text{scalar}_\text{massive} =
  i  \bar \phi D_0 \phi  + \CO(1/m) \ ,
\end{eqnarray}
where we suppress the irrelevant terms.

\subsection{Fermion field}

The Lagrangian for a free massive fermion takes the following form
\begin{eqnarray}
  \CL^\text{fermion} &=&  -i \bar{\Psi}  \gamma^\mu D_\mu \Psi
  \mp i m  \bar{\Psi} \Psi \ .
\end{eqnarray}
Keeping only the particles again, one can expand the fermion field $\Psi$
as
\begin{eqnarray}
  \Psi (t,{\bf x}) =   \Big(
  u_+ \psi_- (t,{\bf x}) + u_- \psi_+ (t,{\bf x}) \Big)
  e^{-im t}\ ,
\end{eqnarray}
where $\psi_\pm$ are single-component Grassmann fields
and $u_\pm$ are orthonormal two-component constant spinors (\ref{constspinor}).
Defining  $D_\pm = D_1\pm  i D_2$ and $A_\pm = A_1\pm i A_2$,
the fermionic Lagrangian can be rewritten as
\begin{eqnarray}
  \CL^\text{fermion} &=& \bar\psi_+
  \Big( i D_0 \psi_-  +m(1\mp 1) \psi_- - iD_- \psi_+  \Big)
  \nonumber \\ &&
  + \bar\psi_- \Big(i D_0 \psi_+ + m (1\pm 1)\psi_+
  - iD_+\psi_- \Big)\ .
\end{eqnarray}
Using the equation of motion for $\bar \psi  $ up to the leading order,
one can show that one of the components $\psi_\pm$ is
completely determined by the other
\begin{eqnarray}
  \left\{ \begin{array}{ll}
  \psi_+ = \frac{i}{2m} D_0 \psi_-  - \frac{i}{2m} D_t \psi_+ \simeq \CO(1/m)
  & \text{for upper sign}\ , \\
  & \\
  \psi_- = \frac{i}{2m} D_- \psi_+ -   \frac{i}{2m} D_t \psi_- \simeq \CO(1/m)
  & \text{for lower  sign}\ . \end{array} \right.
  \label{nonrelf}
\end{eqnarray}
It implies that the spin of dynamical modes in the infinite mass limit
is correlated with the sign of the mass.
Inserting the above relations, the Lagrangian becomes
\begin{eqnarray}
  \CL^\text{fermion}_\text{massive}
  = \left\{ \begin{array}{ll}
  i \bar\psi_+ D_0 \psi_-  + {\cal O}(\frac{1}{m}) & \text{ for upper sign}\ ,
  \\ & \\
  i \bar\psi_-  D_0 \psi_+ + {\cal O}(\frac{1}{m}) & \text{ for lower sign}\ .
  \end{array} \right.
\end{eqnarray}

\subsection{W-boson in Chern-Simons theory}

Let us then discuss the massive W-bosons.
It is well-known that, in the broken phase of
Chern-Simons-matter theories, there is a single massive W-boson that propagates.
We will first discuss how to obtain the polarization vector
for such a propagating mode in the infinite mass limit.

In the broken phase of Chern-Simons-matter theories, the free
Lagrangian for W-boson $W_\mu$ can take the following form
\begin{eqnarray}
  \CL^\text{W-boson} = \pm \frac{k}{2\pi} \e^{\mu\nu\rho}W_\mu^\dagger D_\nu W_\rho
  - v^2 {W_\mu}^\dagger W^\mu\ , \qquad (\e^{012}=1)
\end{eqnarray}
from which one can derive the equation of motions for W-bosons
\begin{eqnarray}
  \frac{k}{2\pi} \e^{\mu\nu\rho} D_\nu W_\rho = \pm v^2 W^\mu\ .
\end{eqnarray}
Here $v$ stands for the vacuum expectation value for the Higgs scalar.
The equation of motion for $\mu=0$ tells us that
the temporal part of vector field behaves like $W_0 \sim \CO(1/v^2)$.
The rest of the equation of motion then reduces to
\begin{eqnarray}
  \e^{ij} D_0 W_j = \mp \frac{2\pi}{k}v^2 W^i + \CO(1/v)\ .
\end{eqnarray}
It implies that the helicity of propagating modes in
the infinite mass limit $v\to\infty$ is again
correlated with the sign of mass as follows:
\begin{eqnarray}
  \vec W \sim\left\{
  \begin{array}{c} \sqrt{\frac{\pi}{k}}
  \vec E_\mp \  w_\pm(t,{\bf x}) e^{-imt} + \CO(1/v)\\
  \sqrt{\frac{\pi}{k}}
  \vec E_\pm  \ w_\mp(t, {\bf x}) e^{+imt} + \CO(1/v)
  \end{array}
  \right. \ ,
\end{eqnarray}
where $\vec E_\pm = ( 1, \pm i)$ denote the polarization
vector with definite helicity $\pm 1$. The Lagrangian of the
particle/anti-particle mode with $k>0$  then becomes
\begin{eqnarray}
  \CL^\text{W-boson}_\text{massive} =
  i \bar w_\mp D_0 w_\pm + \CO(1/m) \ .
\end{eqnarray}

\end{document}